\documentclass{LMCS}

\usepackage{lmodern,amsfonts}
\usepackage{t1enc}
\usepackage{enumerate,hyperref} 

\let \midrule=\hrule
\def \openrule {\vbox \bgroup \halign \bgroup \hfil $##$\hfil \cr}
\def \closerule {\crcr\egroup\egroup}
\def \midclosure {\crcr\noalign {\vskip -0.1em}\closerule}
\let \labclosure = \midclosure

\def \Midclosure {\crcr \noalign
    {\vskip 0.3em \midrule \vskip -0.2em}\midclosure}
\def \Middlerule {\Midclosure\over\openrule}
\def \MIDCLOSURE {\crcr \noalign
    {\vskip 0.3em \midrule \vskip 0.1em \midrule \vskip -0.2em}\midclosure}
\def \MIDDLERULE {\MIDCLOSURE\over\openrule}

\def \lessskip  {\lineskiplimit=.3em\lineskip= .2em plus .1em}
\def \beginrule {\displaystyle \offinterlineskip \lessskip 
    \norulelab \bgroup \openrule}
\def \endrule {\closerule \egroup \rulelab}
\def \mathrule #1{\beginrule #1\endrule}

\let \MIR \MIDDLERULE
\def \setdoublerule {\let \mid \Middlerule \let \labclosure \Midclosure}
\def \settripplerule {\let \mid \MIDDLERULE \let \labclosure \MIDCLOSURE}
\let \nl  \mir

\def \gluelab{{\hskip .5em}}
\def \namelab#1{{\rm(\name{#1})}}

\let \lab = \namelab

\def\rl{\gluelab \namelab}

\def \ll #1{\lab {#1}\gluelab}
\def \Rl #1{\rlap {\rl{#1}}}
\def \Ll #1{\llap {\ll{#1}}}

\def \norulelab {\let \rulelab \relax}
\def \labrule #1{\labclosure \over \gdef \rulelab {#1}\openrule}

\def \rlab #1[#2]{\labrule {\rl {#2}}}
\def \Rlab #1[#2]{\labrule {\ll {#2}}}
\def \llab #1[#2]{\labrule {\Rl {#1}}}
\def \Llab #1[#2]{\labrule {\Ll {#2}}}

\let \and \qquad

\expandafter\ifx\csname name\endcsname\relax
   \def\name#1{\hbox{\sc #1}}\else \relax \fi

\def \beginMathrule #1\nl #2\endMathrule
    {\newdimen\Ruledimen \newdimen\ruledimen
	 \setbox0=\hbox {$\displaystyle \openrule#1\closerule$}\ruledimen=\wd0
	 \setbox0=\hbox {$\displaystyle \openrule#2\closerule$}\Ruledimen=\wd0
	 \ifnum \wd0>\ruledimen
             \advance \Ruledimen by -\ruledimen
             \divide \Ruledimen by 2
           \else
             \Ruledimen=0em
         \fi
        \def \midrule
          {\vskip-0.2em
           \hbox to \ruledimen
                       {\hskip-\Ruledimen\hrulefill\hskip -\Ruledimen}}
	 \beginrule #1\MIR#2\endrule}

\renewcommand{\sum}{\mbox{\sf sum}}
\def\s{{\mbox{\sf s}}}
\def\ts{{\mbox{\tiny\sf s}}}

\def\II{{\mathbb I}}

\def\Pair#1#2#3#4{<#3,#4>_{\Sigma x:#1.#2}}
\def\Pairx#1#2#3#4#5{<#4,#5>_{\Sigma #1:#2.#3}}

\def\implies{\Rightarrow}
\def\exi#1#2#3{\{#1:#2|#3\}}

\def\pair#1#2{\mbox{}<#1,#2> \mbox{}}
\def\qed{\ifmmode\eqno\Box\else\hfill\qquad$\Box$\fi}
\def\prop{\mbox{\sf Prop}}
\def\type{\mbox{\sf Type}}
\def\wf{\vdash \mbox{wf}}
\def\rhde{\rhd_{\beta\varepsilon}}
\def\conv{=_{\beta\varepsilon}}
\def\refl{\mbox{\sf refl}}

\def\eqb{&\equiv &}
\def\I{{\mathcal I}}

\newcommand\bool{\mbox{\sf bool}}
\newcommand\true{\mbox{\sf true}}
\newcommand\false{\mbox{\sf false}}

\newcommand\eqe{\conv}
\newcommand\com{\mbox{\sf com}}
\newcommand\nat{\mbox{\sf nat}}
\newcommand\tab{\mbox{\sf tab}}
\newcommand\acc{\mbox{\sf acc}}
\newcommand\ra{\rightarrow}
\newcommand\la{\leftarrow}
\newcommand\U{{\mathcal U}}

\newcommand\eqrec{\mbox{\sf Eq}\_\mbox{\sf rec}}

\keywords{logic, proofs, coq, lambda-calculus, types}


\def\doi{4 (3:13) 2008}
\lmcsheading%
{\doi}
{1--20}
{}
{}
{Feb.~13, 2007}
{Sep.~26, 2008}
{}   

\begin{document}
\title[Proof-irrelevant type theories]{On the strength of proof-irrelevant type theories}
\author[B.~Werner]{Benjamin Werner}
\address{INRIA Saclay -- Île-de-France and LIX, Ecole Polytechnique, 91128 PALAISEAU cedex, France}
\email{Benjamin.Werner@inria.fr}
\begin{abstract}
We present a type theory with some proof-irrelevance built into the
conversion rule. We argue that this feature is useful
when type theory is used as the logical formalism underlying a theorem
prover. We also show a close relation with the subset types of the
theory of PVS. We show that in these theories, because of the
additional extentionality, the axiom of choice implies the
decidability of equality, that is, almost classical logic. Finally we
describe a simple set-theoretic semantics.
\end{abstract}

\maketitle

\section{Introduction}
\noindent A formal proof system, or proof assistant, implements a formalism in a
similar way a compiler implements a programming language. Among
existing formalisms, dependent type systems are quite widespread. This
can be related to various pleasant features; among them

\begin{enumerate}[(1)]
\item Proofs are objects of the formalism. The syntax is therefore
smoothly uniform, and proofs can be rechecked at will. Also, only the
correctness of the type-checker, a relatively small and
well-identified piece of software, is critical for the reliability of the
system (the ``de Bruijn principle'').

\item\label{conv} The objects of the formalism are programs (typed
$\lambda$-terms) and are identified modulo computation
($\beta$-conversion). This makes the formalism well-adapted for
problems dealing with program correctness. But also the conversion
rule allows the computation steps not to appear in the proof; for
instance $2+2=4$ is simply proved by one reflexivity step, since this
proposition is identified with $4=4$ by conversion. In some cases this
can lead to a dramatic space gain, using the result of certified
computations inside a proof; spectacular recent applications include 
the formal proof of the four-color theorem~\cite{G4} or formal
primality proofs~\cite{GrThWe}. 

\item Finally, type theories are naturally constructive. This makes
stating decidability results much easier. Furthermore, combining this
remark with the two points above, one comes to {\em program
extraction}: taking a proof of a proposition $\forall x:A.\exists
y:B.P(x,y)$, 
one can {\em erase} pieces of the $\lambda$-term in order
to obtain a functional program of type $A\ra B$, whose input and
result are certified to be related by $P$. Up to now however, program
extraction was more an external feature of implemented proof
systems\footnote{Except NuPRL; see related work.}:
programs certified by extraction are no longer objects of the
formalism and cannot be used to assert facts like in the point
above.
\end{enumerate}

\noindent Some related formalisms only build on some of the points above. For
example PVS implements a theory whose objects are functional programs,
but where proofs are not objects of the formalism.

An important remark about~(\ref{conv}) is that the more terms are
identified by the conversion rule, the more powerful this rule is. In
order to identify more terms it thus is tempting to combine
points (2) and (3) by integrating
program extraction into the formalism so that the conversion rule
does not require the {\em computationally irrelevant} parts of terms to be
convertible.

In what follows, we present and argue in favor of a type-theory along
this line. More precisely, we claim that such a feature is useful in at least two
respects. For one, it gives a more comfortable type theory, especially
in the way it handles equality. Furthermore it is a good starting
point to build a platform for programming with dependent types, that
is to use the theorem prover also as a programming
environment. Finally, on a more theoretical level,
we will also see that by making the theory more
extensional, proof-irrelevance brings type theory closer to
set-theory regarding the consequences of the axiom of choice.

The central idea of this work is certainly simple enough to be
adjusted to various kinds of type theories, whether they are
predicative or not, with various kinds of inductive types, more
refined mechanisms to distinguish the computational parts of the
proofs etc\dots. In what follows we illustrate it by using a marking
of the computational content which is as simple as possible.  The
extraction function we use is quite close to Letouzey's
\cite{Letouzey1,Letouzey2}, except that we discard the inclusion
rule $\prop\subset\type$, which would complicate the definition of the
type theory and the semantics (see~\cite{MiqWer} for the last point).

{\bf Related work~} Almost surprisingly, proof-irrelevant type
theories do not seem to enjoy wide use yet. In the literature, they
are often not studied for themselves, but as means for proving
properties of other systems. This is the case for the work of
Altenkirch~\cite{Altenkirch} and Barthe~\cite{Barthe}. One very
interesting work is Pfenning's modal type theory which involves
proof-irrelevance and a sophisticated way to pinpoint which
definitional equality is to be used for each part of a term; in
comparision we here stick to much simpler extraction mechanism.
The NuPRL approach using a squash type~\cite{Caldwell} is very
close to ours, but the extentional setting gives somewhat different
results. Finally, let us mention recent work~\cite{Bruno2} by Barras
and Bernardo who present a type theory with implicit arguments. This
interesting proposal can be understood as a theory with
proof-irrelevance, where the computational fragment is precisely
Miquel's calculus~\cite{miquel}. Their proposal can
be understood as a theory similar to ours, but with a more
sophisticated  way
to mark what is computational and what is not.

\section{The Theory}
\subsection{The \texorpdfstring{$\lambda$}{lambda}-terms}
The core of our theory is a Pure Type System (PTS) extended with $\Sigma$-types
and some inductive type definitions.
In PTS's, the types of types are {\em sorts}; the set of sorts is
$${\mathcal S}\equiv \{ \prop \}\cup \{\type(i) | i\in {\mathbb N} \}$$
As one can see, we keep the sort names of Coq.
As usual, $\prop$ is the impredicative sort and the sorts $\type(i)$
give the hierarchy of predicative universes.  
It comes  as no surprise
that the system contains the usual syntactic constructs of PTSs;
however it is comfortable, both for defining the conversion rule and
constructing a model to {\em tag} the variables to indicate whether
they correspond to a computational piece of code or not; in our case
this means whether  they
live
 in the impredicative or a predicative level (i.e. whether the
type of their type is
$\prop$ or a $\type(i)$). A similar tagging is done on the 
projections of $\Sigma$-types.
Except for this detail, the backbone of the theory
considered hereafter is essentially Luo's Extended Calculus of
Constructions (ECC)~\cite{Luo}. 

The syntax of the ECC fragment is therefore
\def\var#1{#1_{{\ts}}}
\renewcommand{\arraystretch}{1.6}
$$
\begin{array}{ll}
s~~::=~&\prop~|~\type(i) \qquad\qquad\s~~::=~*~|~\diamond\\
t~~::=~&s~|~\var{x}~|~\lambda
\var{x}:t.t~~|~(t~t)~|~\Pi\var{x}:t.t~|~\Sigma^{{\ts}}\var{x}:t.t~
  |~\Pair{t}{t}{t}{t}\\&\phantom{s}~|~\pi_1(t)~|~\pi_2^{\ts}(t)\\
\Gamma~~::=&[]~|~\Gamma(x:t).
\end{array}$$
We sometimes call {\em raw terms} these terms, when we want to stress
that they are considered independently of typing issues.
The tagging of $\Sigma$ is there to indicate whether the second
component of the pair is computational or not (the first component
will always be). For the same technical reason, we also tag the second
projection $\pi_2$.

We will sometimes write $x$ for $x_{\ts}$,
 $\Sigma x:A.B$ for $\Sigma^{{\ts}}
x:A.B$ or $\pi_2(t)$ for $\pi_2^{\ts}(t)$ omitting the tag $\s$ when it
is not relevant or can be infered from the context.

The binding of variables is as usual. We write $t[x\setminus u]$ for
the substitution of the free occurrences of variable $x$ in $t$ by
$u$.  As has become custom, we will not deal with
$\alpha$-conversion here, and leave open the choice between named
variables and de Bruijn indices.

We also use the common practice of writing $A\ra B$ (resp. $A \times B$) for
$\Pi x:A.B$ (resp. $\Sigma x:A.B$)
when $x$ does not appear free in $B$. We also write $\Pi x,y:A.B$
(resp. $\lambda x,y:A.t$) for $\Pi x:A.\Pi y:A.B$ (resp. $\lambda
x:A.\lambda y:A.t$).

\subsection{Relaxed conversion}
The aim of this work is the study of a relaxed conversion rule. While
the idea is to identify terms with respect to typing information, the
tagging of impredicative {\em vs.} predicative variables is sufficient
to define such a conversion in a simple syntactic way. A variable or a
second projection $\pi_2(t)$ is
computationally irrelevant when tagged with the $*$ mark.
This leads
to the following definition.
\def\raeps{\rhd_\varepsilon}

\begin{defi}[Extraction]
We can simply define the extraction relation $\raeps$ as the
contextual closure of the following rewriting equations
\renewcommand{\arraystretch}{1}
$$
\begin{array}{rclrcl}
x_{*} &\raeps& \varepsilon ~~~~~~~~~~~~~~~~&
\lambda x:A.\varepsilon &\raeps & \varepsilon\\
(\varepsilon~t) &\raeps& \varepsilon &
\pi_2^*(t)&\raeps&\varepsilon.
\end{array}
$$
We write $\raeps^*$ for the reflexive-transitive closure of $\raeps$.
We say that a term $t$ is of tag $*$ if $t\raeps^* \varepsilon$ and of
tag $\diamond$ if not. We write $s(t)$ for the tag of $t$.
\end{defi}
\begin{defi}[Reduction]
The $\beta$-reduction $\rhd_\beta$ is defined as the contextual
closure of the following equations
$$
\begin{array}{cl}
(\lambda x^{\ts}:A.t~u) \rhd_\beta t [x^{\ts}\setminus u]~~~~~~~~&\mbox{if $s(u)=\s$}\\
\pi_1(\Pair{A}{B}{a}{b}) \rhd_\beta a~~~~~~~~~&\mbox{if $s(a)=\diamond$}\\
\pi_2^{\ts}(\Pair{A}{B}{a}{b})  \rhd_\beta b~~~~~&\mbox{if $s(b)=\s$}.
\end{array}
$$
\end{defi}
The restrictions on the right-hand side are there in order to ensure
that the tag is preserved by reduction. Without them $(\lambda
x_\diamond:\prop.x_\diamond~\prop)$ can reduce either to $\varepsilon$
or to $\prop$ which would falsify the Church-Rosser property. Actually
we will see that these restrictions are always satisfied on
well-typed terms, but are necessary in order to assert the
meta-theoretic properties below. While these restrictions are specific
to our way of marking computational terms, other methods will probably
yield similar technical difficulties.

The relaxed reduction $\rhde$ is the union of $\rhd_\beta$
and
$\raeps$.  We write $\conv$ for the
reflexive, symmetric
and transitive closure of $\rhde$ and $\rhde^{*}$
 for the transitive-reflexive
closure of $\rhde$.

It is a good feature to have the predicative universes to be embedded
in each other. It has been observed (Pollack, McKinna, Barras\dots)
that a smooth way to present this is to define a syntactic  subtyping
relation which combines this with $=_\beta$ (or here $\conv$). Note
that this notion of subtyping should not be confused with, for
instance, subtyping of subset types in the style of PVS.
\begin{defi}[Syntactic subtyping]
The subtyping relation is defined on raw-terms as the transitive
closure of the following equations
$$\begin{array}{c}
\type(i) \leq \type(i+1)\qquad
T\eqe T' \implies T\leq T'\\
B\leq B' \implies \Pi x:A.B \leq \Pi x:A.B'.
\end{array}$$
\end{defi}

\subsection{Functional fragment typing rules}
The typing rules for the kernel of our theory are given in
PTS-style~\cite{Barendregt} and correspond to Luo's ECC. The
differences are the use of subtyping in the conversion rule and the
tagging of variables when they are ``pushed'' into the context.

The rules are given in figure~\ref{pts}. In the rule {\sc Prod},
$\max$ is the maximum of two sorts for the order $\prop < \type(0) <
\type(1) <\dots$

\begin{figure}
$$
\renewcommand{\arraystretch}{2.77}
\begin{array}{|c|}
\hline
\rl{Prop}
\mathrule{\Gamma\wf\nl
\Gamma\vdash \prop:\type(i)}
\qquad
\rl{Type}
\mathrule{\Gamma\wf\nl
\Gamma\vdash \type(i):\type(i+p)}
\\
\rl{Base}
\mathrule{~\nl
[]\wf}
\qquad
\rl{Var}
\mathrule{
\Gamma\wf \nl
\Gamma\vdash x:A}\mbox{if~}(x:A)\in\Gamma
\\
\rl{Cont}
\mathrule{\Gamma\vdash A:\type(i)\nl
\Gamma(x_\diamond:A)\wf}
\rl{Cont*}
\mathrule{\Gamma\vdash A:\prop\nl
\Gamma(x_{*}:A)\wf}
\\
\rl{Conv}
\mathrule{\Gamma\vdash t:A \qquad
\Gamma\vdash B:s\nl
\Gamma\vdash t:B}\mbox{if~}A\leq  B
\\ 
\rl{Prod}
\mathrule{\Gamma\vdash A: s\qquad
\Gamma(x_{{\ts}}:A)\vdash B:\type(i)\nl
\Gamma\vdash\Pi x_{{\ts}}:A.B : \max(s,\type(i))}
\\
\rl{Prod*}
\mathrule{\Gamma\vdash A:s \qquad
\Gamma(x_{{\ts}}:A)\vdash B:\prop\nl
\Gamma\vdash\Pi x_{\ts} : A . B : \prop}
\\
\rl{Lam}
\mathrule{
\Gamma(x:A)\vdash t:B \nl
\Gamma\vdash \lambda x:A.t : \Pi  x:A.B}
\qquad
\rl{App}
\mathrule{\Gamma\vdash t:\Pi x_{{\ts}}:A.B \qquad
\Gamma\vdash u:A \nl
\Gamma\vdash (t~u) : B[x\setminus u]}
  \ll{\rm if $s(u)=\s$}\\
\rl{Sig}
\mathrule{
\Gamma(x_\diamond:A)\vdash B:\type(i)\nl
\Gamma\vdash \Sigma^\diamond x_\diamond:A.B:\type(i)}
\qquad
\rl{Sig$^*$}
\mathrule{
\Gamma(x_\diamond:A)\vdash B:\prop\nl
\Gamma\vdash \Sigma^* x_\diamond:A.B:\prop}
\qquad
\\
\rl{Pair}
\mathrule{\Gamma\vdash a:A\qquad
\Gamma(x:A)\vdash b:B\qquad
\Gamma\vdash \Sigma^{\ts} x:A.B:\type(i)\nl
\Gamma\vdash \Pair{A}{B}{a}{b}:\Sigma^{\ts} x:A.B}
\\
\rl{Proj1}
\mathrule{\Gamma\vdash t:\Sigma^{\ts} x:A.B\nl
\Gamma\vdash \pi_1 (t) : A}
\qquad
\rl{Proj2}
\mathrule{\Gamma\vdash t:\Sigma^{\ts}  x:A.B\nl
\Gamma\vdash \pi_2^{\ts} (t): B[x\setminus \pi_1(t)]}  \\[7pt]
\hline
\end{array}
$$
\caption{The ECC fragment}\label{pts}
\end{figure}

\subsection{Treatment of propositional equality}\label{eqrec}
Propositional equality is a first example whose treatment changes when
switching to a proof-irrelevant type theory. The definition itself is
unchanged; two objects $a$ and $b$ of a given type $A$ are equal if
and only if they enjoy the same properties
$$a=_A b~~\equiv~~ \Pi P:A\ra\prop.(P~a)\ra(P~b)$$

It is well-known that reflexivity, symmetry and transitivity of
equality can easily be proved. When seen as an inductive definition,
the definition of ``$=_A$'' is viewed as its own elimination
principle.

Let us write $\refl$ for the canonical proof of reflexivity
$$\refl \equiv \lambda A:\type(i).\lambda x:A.\lambda
P:A\ra\prop.\lambda p:(P ~x).p$$

In many cases, it is useful  to extend this elimination over
the computational levels
$$\eqrec_i:\Pi A:\type(i).\Pi P:A\ra\type(i).\Pi a,b:A.(P~a)\ra
a=_A b\ra(P~b)$$
There is however a
peculiarity to $\eqrec$: in Coq, it is defined by case analysis
and therefore comes with a computation rule. The term
$(\eqrec~A~P~a~b~p~e)$ of type $(P~b)$ reduces to $p$ in the case
where $e$ is a canonical proof by reflexivity; in this case, $a$ and
$b$ are convertible and thus coherence and normalization of the type
theory are preserved. 

As shown in the next section, such a reduction rule is useful,
especially when programming with dependent types.  In our
proof-irrelevant theory however, we cannot rely on the information
given by the equality proof $e$, since all equality proofs are treated
as convertible. Furthermore, allowing, for any $e$, the reduction rule
$(\eqrec~A~P~a~b~p~e)\rhd p$ is too permissive, since it easily breaks
the subject reduction property in incoherent contexts.

We therefore put the burden of checking convertibility between $a$ and
$b$ on the reduction rule of $\eqrec$ by extending reduction with the
following,  conditional rule
$$(\eqrec ~A~P~a~b~p~e)\rhd p\hbox{\ \  if } a =_{\varepsilon} b$$
When being precise, this means that $=_{\varepsilon\beta}$ and $\rhd$
are actually two mutual inductive definitions.

An alternative would be the non-linear rule
$$(\eqrec ~A~P~a~a~p~e)\rhd p$$
but this allows an encoding of Klop's counter-example~\cite{Klop} and thus
breaks the Church-Rosser property (for untyped terms). We thus
develop the metatheory for the first version.

\subsection{Generalization}\label{ROC}
In Coq, computational eliminations are provided for more inductive
definitions than just propositional equality. The condition is that
\begin{enumerate}[(1)]
\item The definition has at most one constructor,
\item the arguments of this constructor are all, themselves,
  non-computational.
\end{enumerate}
It appears that it is reasonably straightforward to extend our type theory,
by generalizing the $\eqrec$ feature, in order to capture this Coq
behavior in the case where the inductive definition is non-recursive. We
briefly indicate how but without precise justification. The remainder
of this paragraph is thus not considered in the meta-theoretical
justifications; it is also not necessary for the rest of the article.

We write $\Pi \vec{x}:\vec{A}.T$ for $\Pi x_1:A_1.\dots\Pi x_n:A_n.T$
and $t~\vec{u}$ for $(t~u_1~\dots~u_n)$.

Consider an inductive definition $I: \Pi \vec{x}:\vec{A}.\prop$ with a
unique constructor $c : \Pi \vec{y}: \vec{B}.(I~\vec{u})$. The
non-computational elimination scheme is
$$ \mbox{I\_ind} :  \Pi P : (\Pi \vec{x}:\vec{A}.\prop) .
      (\Pi \vec{y} : \vec{B} . P~\vec{u}) \ra  \Pi\vec{x}:\vec{A}.
         I~\vec{x} \ra  P~\vec{x}$$
with the reduction rule
$$ (\mbox{I\_ind} ~ X ~ p ~\vec{a} ~(c~\vec{b})) \rhd (p~\vec{b}) $$ 
We can then provide a computational elimination
$$\mbox{I\_rec} :  \Pi X : (\Pi \vec{x}:\vec{A}.\type) .
      (\Pi \vec{y} : \vec{B} . X~\vec{u}) \ra  \Pi\vec{x}:\vec{A}.
         I~\vec{x} \ra  X~\vec{x}$$
with the following reduction rule
$$ (\mbox{I\_rec} ~ X ~ p ~\vec{a} ~i) \rhd (p~\vec{\varepsilon})
\qquad \mbox{if~}\vec{u}=_{\beta\varepsilon}\vec{a}$$
To understand the last condition, one should note that although the
variables $\vec{y}$ are free in $\vec{u}$, they do not interfere with
the conversion since their types ensure they are all tagged by $*$.

\subsection{Data Types}
In order to be practical, the theory needs to be  extended by inductive
definitions in the style of Coq, Lego and others. 
We do not detail the typing rules and liberally use integers, booleans,
usual functions and predicates ranging over them. We refer to
the Coq documentation~\cite{Coq,Gim}; for a possibly more modern
presentation~\cite{Blanqui} is interesting.

Let us just mention that data types live in \type. That is, for instance,
  $\nat:\type(0)$; thus, their elements are of tag $\diamond$.

\section{Basic metatheory}

\noindent We sketch the basic meta-theory of the calculus defined up to
here.  The proof techniques are relatively traditional, even if
one has to take care of the more delicate behavior of relaxed
reduction for the first lemmas (similarly to~\cite{MiqWer}).

\begin{lem}\label{preserv}
If $t \rhd_\beta t'$, then $s(t)=s(t')$. Thus, the same is true if
$t\rhd^*_{\beta\varepsilon} t'$.
\end{lem}
\begin{proof}
By a straightforward case analysis of the form of $t$.
\end{proof}

\begin{lem}[$\beta$-postponement] If 
  $t\rhde^* t'$, then there exists $t''$ such that $t\raeps^*
  t''$ and $t''\rhd_\beta^* t'$.
\end{lem}
\begin{proof}
One first shows that if $t\rhd_\beta t' \raeps t''$, then
there exists $t'''$ such that $t\raeps^* t'''$ and either
$t'''=t''$ or $t'''\rhd_\beta t''$. This is done by checking how the
two redexes are located with respect to each other.
The proof of the lemma then easily follows.
\end{proof}

\begin{lem}[Church-Rosser]
For $t$ a raw term, if $t\rhde^* t_1$ and $t\rhde^* t_2$,
then there exists $t_3$ such that $t_2\rhde^* t_3$ and $t_2\rhde^*
t_3$.
\end{lem}
\begin{proof}
By a quite straightforward adaptation of the usual Tait and
Martin-L{\"o}f method. The delicate point was to choose the right
formulation of the reduction rule specific to the elimination of
propositional equality, as mentioned in section~\ref{eqrec}.
\end{proof}
An immediate but very important consequence is that
\begin{cor}[Uniqueness of product formation]\label{UTF}
If $\Pi x : A.B \leq \Pi x : A'. B'$, then $A \conv A'$ and $B\leq
B'$.
\end{cor}
\begin{cor}\label{UTFF}
For any $T$,
\begin{enumerate}[$\bullet$]
\item $T\leq \prop \iff \prop\leq T \iff T\conv\prop \iff T \rhde^*\prop $
\item $ T\leq \type(i) \iff T \rhde^* \type(j) ~\mbox{with}~j\leq i$
\item $ \type(i) \leq T \iff T \rhde^* \type(j) ~\mbox{with}~i\leq j$
\item if $T\leq U$ and $U\leq T$ then $U\conv T$.
\end{enumerate}
\end{cor}

Furthermore, $\raeps$ is obviously strongly normalizing. One
therefore can "pre-cook" all terms by $\raeps$ when checking
relaxed convertibility.
\begin{lem}[pre-cooking of terms]\label{cook}
Let $t_1$ and $t_2$ be raw terms. Let $t_1'$ and $t_2'$ be
their respective $\raeps$-normal forms. Then, $t_1\eqe t_2$
if and only if $t_1'=_\beta t_2'$.
\end{lem}
While this property is important for implementation, its converse is
also true and semantically understandable. Computationally relevant
$\beta$-reductions are never blocked by not-yet-performed
$\varepsilon$-reductions.
\begin{lem}\label{uncook}
Let $t_1$ be any raw term. Suppose $t_1\raeps t_2 \rhd_\beta
t_3$. Then there exists $t_4$ such that $t_1\rhd_\beta
t_4\raeps^*t_3$.
\end{lem}
\begin{proof}
It is easy to see that $\raeps$ cannot create new
$\beta$-redexes, nor does it duplicate existing ones.
\end{proof}

\begin{lem}
If $t\rhde t'$, for any term $u$ and variable $x_{s(u)}$, one has
$t[x_{s(u)}\setminus u] \rhde t'[x_{s(u)}\setminus u]$. Thus, if
$t\conv t'$ then $t[x_{s(u)}\setminus u] \conv t'[x_{s(u)}\setminus
u]$.
\end{lem}
\begin{proof}
By straightforward induction over the structure of $t$. One uses the
fact that, since $x_{s(u)}$ and $u$ have the same syntactic sort, the
terms  $t$ and $t[x_{s(u)}\setminus u]$ also have the same syntactic
sort.
\end{proof}

\begin{lem}[Substitution]\label{wtsubst}
If $\Gamma(x:A)\Delta\vdash t:T$ and $\Gamma\vdash a:A$ are derivable,
if $a$ and $x$ have the same (syntactic) sort, 
then
$\Gamma\Delta[x\setminus a] \vdash t[x\setminus a]:T[x\setminus a]$ is
derivable.
\end{lem}
\begin{proof}
By induction over the structure of the first derivation, like in the
usual proof. The condition over the syntactic sorts is necessary for
the case of the conversion rule, in order to apply the previous lemma.
\end{proof}
\smallskip

\begin{lem}[Inversion or Stripping]~If $\Gamma\vdash t:T$ is derivable, then so are $\Gamma\wf$ and
$\Gamma\vdash T:s$ for some sort $s$. Furthermore, the following
clauses hold.

~

\noindent
\begin{tabular}{|l|l|}\hline
If $\Gamma\vdash x : T$ is derivable, then:
&
If $\Gamma\vdash (t~u):V$ is derivable, then:\\
\begin{tabular}{l}
$\bullet$~ $(x,U)\in\Gamma$,\\
$\bullet$~ $\Gamma\vdash T : s$ is derivable,\\
$\bullet$~ $U\leq T$.
\end{tabular}
&
\begin{tabular}{l}
$\bullet$~ $\Gamma\vdash t : \Pi x : U.W$,\\
$\bullet$~ $\Gamma\vdash u : U$,\\
$\bullet$~ $W[x\setminus u]\leq V$.
\end{tabular}
\\ \hline
If $\Gamma\vdash \lambda x : U.t : W$ is derivable, then: &

If $\Gamma\vdash \Pi x : A . B:T$ is derivable,
then \\
\begin{tabular}{l}
$\bullet$~ $\Gamma(x:U)\vdash t : T$,\\
$\bullet$~ $\Pi x : U. T\leq W$.\\
~\\
~
\end{tabular} &
\begin{tabular}{l}
$\bullet$~ $\Gamma\vdash A : s_1$,\\
$\bullet$~ $\Gamma(x:A)\vdash B : s_2$,\\
$\bullet$~ either $s_2=\prop$ and $\prop\leq T$\\
 or $max(s_1,s_2)\leq T$.
\end{tabular}\\  \hline
If $\Gamma\vdash \Sigma^* x_\diamond: A . B:T$ is derivable,
then &
If $\Gamma\vdash \Sigma^\diamond x_\diamond: A . B:T$ is derivable,
then \\
\begin{tabular}{l}
$\bullet$~ $\Gamma\vdash A : \type(i)$,\\
$\bullet$~ $\Gamma(x:A)\vdash B : \prop $,\\
$\bullet$~  $\type(i) \leq T$.
\end{tabular}&
\begin{tabular}{l}
$\bullet$~ $\Gamma\vdash A : \type(i)$,\\
$\bullet$~ $\Gamma(x:A)\vdash B : \type(j) $,\\
$\bullet$~  $\type(max(i,j)) \leq T$.
\end{tabular}
\\ \hline
$\Gamma\vdash \Sigma x_*:A.B:T$ is not derivable. &
If $\Gamma\vdash \Pairx{x_\diamond}{T}{U}{t}{u} : V$ is derivable,\\
\begin{tabular}{l}
~\\
~\\
~\\
~
\end{tabular}&
\begin{tabular}{l}
$\bullet$~ $\Sigma x_\diamond:T.U\leq V$,\\
$\bullet$~ $\Gamma\vdash t:T$,\\
$\bullet$~ $\Gamma\vdash u : U[x_\diamond\setminus t]$,\\
$\bullet$~ $s(t)=\diamond$.
\end{tabular}\\ \hline
If $\Gamma\vdash \pi_1(t) : T$ is derivable, then &
If  $\Gamma\vdash \pi_2 ^{{\ts}}(t) : T$ is derivable, then
\\
\begin{tabular}{l}
$\bullet$~ $\Gamma\vdash t:\Sigma x_\diamond:A.B$,\\
$\bullet$~ $A \leq T$.
\end{tabular}&
\begin{tabular}{l}
$\bullet$~ $\Gamma\vdash t:\Sigma^{{\ts}} x_\diamond:A.B$,\\
$\bullet$~ $ B[x_\diamond \setminus \pi_1(t)] \leq T$
\end{tabular}\\ \hline
If $\Gamma\vdash \prop : T$ is derivable, then $\type(1) \leq T$
&
If $\Gamma\vdash \type(i): T$, then $\type(i+1)\leq T$\\ \hline
\end{tabular}
\end{lem}
\begin{proof}
Simultaneously by induction over the derivation.
\end{proof}
\begin{cor}[Principal type]
If $\Gamma\vdash t:T$, then there exists $U$ such that $\Gamma\vdash
t:U$ and for all $V$, if $\Gamma\vdash t:V$, then  $U \leq V$.
\end{cor}
\begin{proof}
By induction over the structure of $t$, using the previous lemma and
corollaries~\ref{UTF} and~\ref{UTFF}.
\end{proof}

Of course, subject reduction holds only for  $\rhd_\beta$-reduction,
since $\varepsilon$ is not meant to be typable.
\begin{lem}[Subject reduction]
If $\Gamma\vdash t:T$ is derivable, if $t\rhd_\beta t'$ (resp. $T\rhd_\beta T'$,
$\Gamma\rhd_\beta \Gamma'$), then
$\Gamma\vdash t':T$ (resp. $\Gamma\vdash t:T', \Gamma'\vdash t:T$).
\end{lem}
\begin{proof}
By induction over the structure of $t$. Depending upon the position of
the redex, one uses either the substitution or the stripping lemmas
above. We only detail the case where a $\beta$-reduction occurs at the
root of the term.

If $t=\lambda x^{\ts} : U.v~u$, $s(u)={\ts}$ and $t'=v[x^{\ts}\setminus u]$, we
know that $\Gamma(x^{\ts}:U)\vdash v:V$, $\Gamma\vdash u:U$ and
$V[x^{\ts}\setminus u]\leq T$. Thus we can apply lemma~\ref{wtsubst} to
deduce
$$\Gamma\vdash v[x^{\ts}\setminus u]:V[x^{\ts}\setminus u]$$
and
$$\Gamma\vdash V[x^{\ts}\setminus u]: s$$
where $s$ is the sort such that $\Gamma(x^{\ts}:U)\vdash V:s$. The result
then follows through one application of the conversion rule.
\end{proof}

\begin{lem}
If $\Gamma\vdash t:T$ is derivable, then there exists a sort $s$ such
that $\Gamma\vdash T:s$; furthermore $\Gamma\vdash T:\prop$ if and
only if $t$ is of tag $*$.
\end{lem}
\begin{proof}
By induction over the structure of $t$. The Church-Rosser property
ensures that $\prop$ and $\type(i)$ are not convertible.
\end{proof}

A most important property is of course  normalization. We do not
claim any proof here, although we very strongly conjecture it. A
smooth way to prove it is probably to build on top of the simple
set-theoretical model using an interpretation
of types as saturated $\Lambda$-sets as first proposed by
Altenkirch~\cite{Alti,MelWer}.

\begin{conj}[Strong Normalization]
If $\Gamma\vdash t:T$ is derivable, then $t$ is strongly normalizing.
\end{conj}
Stating strong normalization is important in the practice of
proof-checking, since it entails decidability of type-checking and
type-inference.
\begin{cor}
Given $\Gamma$, it is decidable whether $\Gamma\wf$. Given $\Gamma$
and a raw term $t$, it is decidable whether there exists $T$ such that
$\Gamma\vdash t:T$ holds.
\end{cor}
\begin{proof}
By induction over the structure of $t$, using the stripping
lemma. Normalization ensures that the relation $\leq$ is decidable for
well-formed types.
\end{proof}

The other usual side-product of normalization is a syntactic assessment
of constructivity.
\begin{cor}\label{constr}
If $[]\vdash t:\Sigma x:A.B$, then $t\rhd_\beta^*<a,b>_{\Sigma x:A.B}$
with $[]\vdash a:A$ and $[]\vdash b:B[x\setminus a]$.
\end{cor}
\begin{proof}
By case analysis over the normal form of $t$, using the stripping
lemma.
\end{proof}

\section{Programming with dependent types}
\noindent We now list some applications of the relaxed conversion rule, which
all follow the slogan that proof-irrelevance makes programming with
dependent types more convenient and efficient.

From now on, we will write $\{x:A|P\}$ for $\Sigma^*x:A.P$, that is
for a 
$\Sigma$-type whose second component is non-computational.

\subsection{Dependent equality}
Programming with dependent types means that terms occur in the type of
computational objects (i.e. not only in propositions). The way
equality is handled over such families of types is thus a crucial
point which is often problematic in intensional type theories.

Let us take a simple example. Suppose we have defined a data-type of
arrays over some type $A$. If $n$ is a natural number, $(\tab~n)$ is
the type of arrays of size $n$. That is $\tab : \nat \ra \type(i)$. 
Furthermore, let us assume we have a function modeling access to the
array
$\acc : \Pi n:\nat. \tab ~n\ra\nat\ra A$.

Commutativity of addition can be proved in the theory: $\com:\Pi
m,p:\nat.(m+p)=(p+m)$. Yet $\tab~(m+p)$ and $\tab~(p+m)$ are two
distinct types with distinct inhabitants. For instance, if we have an
array $t:\tab~(m+p)$, we can use the operator $\eqrec$ described above
to transform it into an array of size $p+m$
$$t'\equiv \eqrec~nat~\tab~(m+p)~(p+m)~t~(\com~(m+p)~(p+m)) ~:~ \tab (p+m)$$
Of course, $t$ and $t'$ should have the same inhabitants, and we would
like to prove
$$\Pi i:\nat.\acc ~(m+p)~t~i =_A \acc~(p+m)~t'~i$$
It is known~\cite{HofStr,McBride} that in order to do so, one needs
the reduction rule for $\eqrec$ together with a proof that equality
proofs are unique. The latter property being generally established by
a variant of what Streicher calls the ``K axiom''
$$K:\Pi A:\type.\Pi a:A.\Pi P:a=_A a\ra\prop.(P~(\refl~a))\ra\Pi
e:a=_A a.(P~e)$$
where $\refl$ stands for the canonical proof by reflexivity. 

Here, since equality proofs are also irrelevant to conversion, this
axiom becomes trivial. Actually, since $(P~e)$ and $(P~(\refl~a))$ are
convertible, this statement does not even need to be mentioned
anymore, and the associated reduction rule becomes superfluous.

In general, it should be interesting to
transpose work like McBride's~\cite{McBride} in the framework of
proof-irrelevant theories.

\subsection{Partial functions and equality over subset types}

In the literature of type theory, subset types come in many flavors;
they designate the restriction of a type to the elements verifying a
certain predicate.  The type $\exi{x}{A}{P}$ can be viewed as the
constructive statement "there exists an element of $A$ verifying $P$",
but also as the data-type $A$ restricted to elements verifying $P$. In
most current intensional type theories, the latter approach is not
very practical since equality is defined over it in a too narrow
way. We have $\pair{a}{p}=_\beta\pair{a'}{p'}$ only if $a=_\beta a'$
{\em and} $p=_\beta p'$; the problem is that one would like to get rid
of the second condition. The same is true for propositional Leibniz
equality and one can establish
$$\pair{a}{p}=_{\exi{x}{A}{P}}\pair{a}{p'}\ra p=_{P[x\setminus a]}p'$$
In general however, one is only interested in the validity of the
assertion $(P~a)$, not the way it is proved. A program awaiting an
argument of type $\exi{x}{A}{P}$ will behave identically if fed with
$\pair{a}{p}$ or $\pair{a}{p'}$. 

Therefore, each time a construct $\exi{x}{A}{P}$ is used indeed as a
data-type, one cannot use Leibniz equality in practice. Instead, one
has to define a less restrictive equivalence relation $\simeq_{A,P}$
which simply states that the two first components of the pair are
equal
$$\pair{a}{p}\simeq_{A,P}\pair{a'}{p'}~~~~\equiv~~~~a=_A a'$$
But using $\simeq_{A,P}$ instead of $=_{\exi{ x}{A}{P}}$ quickly becomes
very tedious;
typically, for every function $f:\exi{x}{A}{P}\ra B$ one has to prove
$$\Pi c,c':\exi{x}{A}{P}~.~c\simeq_{A,P}c'\ra(f~c)=_B(f~c')$$
and even more specific statements if $B$ is itself a subset type.

In our theory, one can prove without difficulties that $=_{\exi{x}{A}{P}}$ and
$\simeq_{A,P}$ are equivalent, and there is indeed no need anymore for
defining $\simeq_{A,P}$. Furthermore, one has
$\pair{a}{p}\eqe\pair{a}{p'}$, so the two terms are computationally
identified which is stronger than Leibniz
equality, avoids the use of the deductive level and makes proofs and
developments more concise.

\subsubsection*{Array bounds}
The same can be observed when partial functions are curryfied. Let us
take again the example of arrays, but suppose this time the access
function awaits a proof that the index is within the bounds of the
array.
\begin{eqnarray*}
\tab &:&\nat\ra\type(i)\\
\acc&:&\Pi n:nat.\tab~n\ra \Pi i:nat.i<n\ra A
\end{eqnarray*}
So given an array $t$ of size $n$, its corresponding access function
is
$$a\equiv \acc~n~t ~:~\Pi i:nat.i<n\ra A$$
In traditional type theory, this definition is cumbersome to use,
since one has to state explicitly that the values  $(a~i~p_i)$, where
$p_i:i<n$ {\em do not depend upon $p_i$}. The type above is therefore
not sufficient to describe an array; instead one needs the additional
condition
$$T_{irr} : \Pi i:nat.\Pi p_i,p_i':i<n.(a~i~p_i)=_A (a~i~p_i')$$
where $=_A$ stands for the propositional Leibniz equality.

This is again verbose and cumbersome since $T_{irr}$ has to be invoked
repeatedly. In our theory, not only the condition $T_{irr}$ becomes
trivial, since for
any $p_i$ and $p_i'$ one has $(a~i~p_i)\eqe (a~i~p_i')$, but this last
coercion is stronger than propositional equality: there is no need
anymore to have recourse to the deductive level and prove this
equality. The proof terms are therefore clearer and smaller.

\subsection{On-the-fly extraction}
An important point, which we only briefly mention here is the
consequence for the implementation when switching to a
proof-irrelevant theory. In a proof-checker, the environment 
consists of a sequence of definitions or lemmas which have been
type-checked. If the proof-checker implements a proof-irrelevant
theory, it is reasonable to keep two versions of each constant: the
full proof-term, which can be printed or re-checked, and the extracted
one (that is $\raeps$-normalized) which is used for conversion
check. This would be even more natural when building on recent Coq
implementations which already use a dual storing of constants, the
second representation being non-printable compiled code precisely used
for fast conversion check.

In other words, a proof-system built upon a theory as the one
presented here would allow the user to efficiently exploit the
computational behavior of a constructive proof in order to prove new
facts. This makes the benefits of program extraction technology
available inside the system and helps transforming proof-system into
viable programming environments.

\section{Relating to PVS}\label{pvs}
\noindent Subset types also form the core of PVS. In this formalism the objects
of type $\{x:A | P \}$ are also of type $A$, and objects of type $A$
can be of type $\{x:A | P \}$. This makes type checking
undecidable and is thus
impossible in our setting. But we show that it is possible to build explicit
coercions between the corresponding types of our theory which
basically behave like the identity.

What is presented in this section is strongly related to the work of
Sozeau~\cite{sozeau}, which describes a way to provide a PVS style
input mode for Coq.

The following lemma states that the construction and destruction
operations of our subset types can actually be omitted when checking
conversion.
\begin{lem}[Singleton simplification]
  The typing relation of our theory remains unchanged if we extend the
  $\raeps$ reduction of our theory by\footnote{To make the second
  clause rigorous, a solution is to modify slightly the theory by
  adding a tag the first projection ($\pi_1 ^*(t)$ and
  $\pi_1^{\diamond}(t)$). This does not significantly change the
  metatheory.}.
\begin{eqnarray*}
\pair{a}{p}_{\Sigma^*x:A.P} &\raeps& a\\
\pi_1(c) &\raeps& c ~~~\mbox{when~$c:\Sigma^* x:A.B$}
\end{eqnarray*}
\end{lem}

The following definition is directly 
transposed\footnote{A difference
is that in PVS, propositions and booleans are identified; but this
point is independent of this study. It is however possible to do the
same in our theory  by assuming a computational version of
excluded-middle.}
 from
PVS~\cite{pvs}. We do not treat dependent types in full generality
(see~chapter 3 of~\cite{pvs}).
\begin{defi}[Maximal super-type]
The maximal super-type is a partial function $\mu$ from terms to terms,
recursively defined by the following equations. In all these
equations, $A$ and $B$ are of type $\type(i)$ in a given context.
$$
\begin{array}{rcllrcl}
\mu(A)\eqb A&\mbox{\hskip -1.3 cm if $A$ is a data-type}~~~~~~~~~&\hskip 1 cm
~~~~\mu(\exi{x}{A}{P})\eqb\mu(A)\\
\mu(A\ra B)\eqb A\ra\mu(B)&&
\mu(A\times B)\eqb \mu(A)\times\mu(B).
\end{array}$$
\end{defi}

\begin{defi}[$\eta$-reduction]
The generalized $\eta$-reduction, written $\rhd_\eta$, is the
contextual closure of
\begin{eqnarray*}
\lambda x:A.(t~x)&\rhd_\eta&t~\mbox{~~~~~if $x$ is not free in $t$}\\
<\pi_1(t),\pi_2(t)>&\rhd_\eta& t
\end{eqnarray*}
\end{defi}
We can now construct the coercion function from $A$ to $\mu(A)$.
\begin{lem}
If $\Gamma\vdash A:\type(i)$ and $\mu(A)$ is defined, then
\begin{enumerate}[$\bullet$]
\item $\Gamma\vdash \mu(A):\type(i)$,
\item there exists a function $\overline{\mu}(A)$ which is of type
$A\ra\mu(A)$ in $\Gamma$,
\item furthermore, when applying the singleton simplification ${\mathcal
S}$ to $\overline{\mu}(A)$ one obtains an $\eta$-expansion of the
identity function; to be precise,
${\mathcal S}(\overline{\mu}(A))\rhd_{\varepsilon\beta\eta}^* \lambda x:A.x.$
\end{enumerate}
\end{lem}
\proof
It is almost trivial to check that $\Gamma\vdash\mu(A):\type(i)$. The
two other clauses are proved by induction over the structure of
$A$.
\begin{enumerate}[$\bullet$]
\item If $A$ is of the form $\{x:B | P \}$ with
  $\overline{\mu}(B):B\ra\mu(B)$, then
\[\overline{\mu}(A)\equiv\lambda
  x:\exi{x}{B}{P}.(\overline{\mu}(B)~\pi_1(x)):\exi{x}{B}{P}\ra\mu(B)
\]
  Furthermore,
  since $P:\prop$,
  $\pi_1(x)$ is here simplified to $x$, and by induction hypothesis we
  know that 
  ${\mathcal S}(\overline{\mu}(B))~x$ reduces to $x$. We can conclude
  that ${\mathcal S}(\overline{\mu}(A))\rhd_{\varepsilon\beta\eta}^*
  \lambda x:\exi{x}{B}{P}.x$.
\item If $A$ is of the form $C\ra B$ with $\overline{\mu}(B):B\ra\mu(B)$, then
\[\overline{\mu}(A)\equiv\lambda
  h: C\ra B.\lambda x : C.\overline{\mu}(B)~(h~x) : C\ra\mu(B)
\]
Since $({\mathcal S}(\overline{\mu}(B))~(h~x))\rhd_{\varepsilon\beta\eta}^*
(h~x)$, we have ${\mathcal S}(\overline{\mu}(A))\rhd_{\varepsilon\beta\eta}^*\lambda
h:A\ra B.h$.
\item If $A$ is of the form $B\times C$, then
\[\overline{\mu}(A)\equiv \lambda
x:B\times
C.<(\overline{\mu}(B)~\pi_1(x)),(\overline{\mu}(C)~\pi_2(x))>_{\mu(B)\times
  \mu(C)}
\]
 Again, the induction hypotheses assure that
$\overline{\mu}(A)\rhd_{\varepsilon\beta\eta}^*\lambda x:B\times C.x$.\qed
\end{enumerate}

The opposite operation, going from from $\mu(A)$ to $A$, can only be
performed when some conditions are verified ({\em type-checking
  conditions}, or TCC's in PVS
terminology). We can also transpose this to our theory, still keeping the
simple computational behavior of the coercion function. This time
however, our typing being less flexible than PVS', we have to define the
coercion function and its type simultaneously; furthermore, in
general, this operation is well-typed only if the type-theory supports
generalized
$\eta$-reduction\footnote{It should be mentionned that
  adding $\eta$-reduction to such a type system yields non-trivial
  technical difficulties, which are mostly independent of the question
  of proof-irrelevance.}.

This unfortunate restriction is typical when defining transformations
over programs with dependent types. It should however not be taken too
seriously, and we believe this cosmetic imperfection can generally be
tackled in practice\footnote{For one, in practical cases, $\eta$-does
not seem necessary very often (only with some nested
existentials). And even then, it should be possible to tackle the
problem by proving the corresponding equality on the deductive level.}.
\begin{lem}[subtype constraint]
Given $\Gamma\vdash A:\type(i)$, if $\mu(A)$ is defined, then one can
define $\pi(A)$ and $\overline{\pi}(A)$ such that, {\em in the theory
where conversion is extended with $\rhd_\eta$}, one has
$$\Gamma\vdash \pi(A) : \mu(A)\ra\prop~~~~\mbox{and}~~~
\Gamma\vdash \overline{\pi}(A) : \Pi x:\mu(A).(\pi(A)~x)\ra A$$
Furthermore, $\overline{\pi}(A)$ $\rhd_{\varepsilon\beta\eta}$-normalizes
to $\lambda x:\mu(A).\lambda p:(\pi(A)~x).x$.
\end{lem}
\begin{proof}
By straightforward induction. We only provide detail for the case where
$A=B\ra C$. Then $\pi(A)\equiv\lambda f:A\ra\mu(B).\forall
x:A.(\pi(B)~(f~x))$ and $\overline{\pi}(A)\equiv\lambda
f:A\ra\mu(B).\lambda p :\forall x:A.(\pi(B)~(f~x)).\lambda x:A.
(\overline{\pi}(B)~(f~x)~(p~x))$.
\end{proof}
\vfill\eject

\section{A more extensional theory}
\noindent Especially during the 1970s and 1980s, there was an intense
debate about the respective advantages of intensional versus
extensional type theories. The latter denomination seems to cover
various features like replacing conversion by propositional equality
in the conversion rule or adding primitive quotient types. In general,
these features provide a more comfortable construction of some
mathematical concepts and are closer to set-theoretical practice. But
they break other desirable properties, like decidability of
type-checking and strong normalization.

The theory presented here should therefore be considered as belonging
to the intentional family. However, we retrieve some features usually
understood as extensional.

\subsection{The axiom of choice}
Consider the usual form of the (typed) axiom of choice (AC)
$$(\forall x:A. \exists y:B . R(x,y)) \implies \exists f:A\ra
B. \forall x:A . R(x,f~x)$$

When we transpose it into our type theory, we can choose to translate
the existential quantifier either by a $\Sigma$-type, or the
existential quantifier defined in $\prop$ 
$$\exists x:A.P \equiv \Pi Q:\prop . (\Pi x:A . P\ra Q) \ra Q ~: \prop$$

If we use a $\Sigma$-type, we get a type which obviously inhabited,
using the projections $\pi_1$ and $\pi_2$. However, if we read the
existential quantifiers of AC as defined above, we obtain a
(non-computational) proposition which is not provable in type theory.

Schematically, this propositions states that if $\Pi x:A.\exists
y:B.R(x,y)$
is provable, then the corresponding function from $A$ to
$B$ exists ``in the model''. This assumption is strong and allows to encode IZF
set theory into type theory (see~\cite{Werner}). 

What is new is that our proof-irrelevant  type theory is extensional
enough to perform the first part of Goodman and Myhill's proof based
on Diaconescu's observation. Assuming AC, we can prove the
decidability of equality. Consider any type $A$ and two objects $a$ and
$b$ of type $A$. We define a type corresponding to the unordered pair
$$\{a,b\}\equiv \exi{x}{A}{ x=_A a\vee x=_A b}$$
Let us write $a'$ (resp. $b'$) for the element of $\{a,b\}$ corresponding
to $a$ (resp. $b$); so $\pi_1(a')\conv a$ and
$\pi_1(b')\conv b$.
It is then easy to prove that 
$$\Pi z:\{a,b\} . \exists e:\bool . (e=_{\bool}\true\wedge \pi_1(z)=_A
a)\vee(e=_{\bool}\false\wedge \pi_1(z)=_A b)$$
and from the axiom of choice we deduce
$$\exists f : \{a,b\}\ra \bool . \Pi z:\{a,b\} . (f~z=_{\bool}\true
\wedge \pi_1(z)=_A a)\vee(f~z=_{\bool}\false \wedge \pi_1(z)=_A b)$$
Finally given such a function $f$, one can compare $(f~a')$ and
$(f~b')$, since both are booleans over which equality is decidable.

The key point is then that, thanks to proof-irrelevance, the
equivalence between $a'=_{\{a,b\} }b'$ and $a=_A b$ is provable in the theory.
Therefore, if $(f~a')$ and $(f~b')$ are different, so are $a$ and
$b$. On the other hand, if $(f~a')=_{\bool}(f~b')=_{\bool}\true$ then
$\pi_1(b')=_A a$
and so $b=_A a$. In the same way, $(f~a')=_{\bool}(f~b')=_{\bool}\false$ entails $b =_A a$.

We thus deduce $a=_A b \vee a \not =_A b$ and by generalizing with respect to
$a, b$ and $A$ we obtain
$$\Pi A:\type(i).\Pi a,b : A . a=_A b \vee a \not=_A b$$
which is a quite classical statement. We have formalized this proof in
Coq, assuming proof-irrelevance as an axiom.

Note of course that this ``decidability'' is restricted to a
disjunction in $\prop$ and that it is not possible to build an actual
generic decision function. Indeed, constructivity of results in the
predicative fragment of the theory are preserved, even if assuming the
excluded-middle in $\prop$.

\subsection{Other classical non-computational axioms}
At present, we have not been able to deduce the excluded middle (EM) from
the statement above\footnote{In set theory, decidability of equality
  entails the excluded middle, since $\{x\in\mathbb N | P \}$ is equal
  to $\mathbb N$ if and only if $P$ holds.}. We leave this theoretical
question to future investigations but it seems quite clear that in most
cases, when admitting AC one will also be willing to admit EM. In fact
both axioms are validated by the simple set-theoretical model and give
a setting where the $\type(i)$'s are inhabited by computational types
(i.e. from $\exi{x}{A}{P}$ we can compute $x$ of type $A$) and $\prop$
allows classical reasoning about those programs.

Another practical statement which is validated by the set-theoretical
model is the axiom that point-wise equal functions are equal
$$ \mbox{(EXT)}~~~~~\Pi A, B : \type(i) . \Pi f,g : A\ra B . (\Pi x : A . f~x =_B g~x) \ra
f=_{A\ra B}g$$
Note that combining this axiom with AC (and thus decidability of
equality) is already enough to prove (in $\prop$) the existence of a
function deciding whether a Turing machine halts.

\subsection{Quotients and normalized types}

Quotient sets are a typically extensional concept whose adaptation to
type theory has always been problematic. Again, one has to choose
between ``effective'' quotients and decidability of type-checking.
Searching for a possible compromise, Courtieu~\cite{Courtieu} ended up
with an interesting notion of {\em normalized type}\footnote{A similar
  notion has been developed for NuPRL~\cite{NK06}.}. The idea is
remarkably simple: given a function $f:A\ra B$, we can define 
$\{(f~x) | x: A\}$ which is the subtype of $B$ corresponding to the
range of $f$. His rules are straightforwardly translated into our
theory  by simply taking
$$\{f(x)|x:A\} \equiv \exi{y}{B}{ \exists x:A . y=_B f~x}$$

Courtieu also gives the typing rules for functions going from $A$ to
$\{f(x)|x:A\}$, and back in the case where $f$ is actually of type
$A\ra A$. 

The relation with quotients being that in the case $f:A \ra A$ we can
understand $\{f(x)|x:A\}$ as the type $A$ quotiented by the relation 
$$ x~R~y ~~\iff~~f~x=_A f~y$$

In practice this appears to be often the case, and Courtieu describes
several applications.
\vfill\eject

\section{Simple semantics}
\label{sem}
\noindent When justifying the correctness of a program extraction
mechanism, one can use either semantics or syntax. In the first case,
one builds a model and verifies it validates
extraction~\cite{Berardi}. In the latter case, at least in the
framework of type theories, this mainly means building a realizability
interpretation on top of the strong normalization
property~\cite{Paulin}. This second approach is difficult here, since
our theory is {\em itself} built using the erasure of
non-computational terms. Furthermore, for complex theories, it appears
easier to prove strong normalization using an already defined
model~\cite{Alti,MelWer,CoqSpi}.

For this reason alone, it is worth treating the topic of semantics
here. Furthermore, we believe it is a good point for a theory meant to
be used in a proof-system to bear simple semantics, in order to
justify easily the validity of additional axioms like the ones
mentioned in the previous section or extensions like the useful
reduction rule for $\eqrec$ (par.~\ref{eqrec}) which is difficult to
treat by purely syntactic means.

Set-theoretical interpretations are the most straightforward way to
provide semantics for typed $\lambda$-calculi. It consists, given an
interpretation $\mathcal I$ of the free variables, of
interpreting a type $T$ by a set $|T|_{\mathcal I}$, and terms $t:T$ by
elements $|t|_{\mathcal I}$ of $|T|_{\mathcal I}$. Furthermore,
$\lambda$-abstractions are interpreted by their set-theoretical
counterparts: $|\lambda x:A.t|_{\mathcal I}$ is the function mapping
$\alpha\in|A|_{\mathcal I}$ to $|t|_{\mathcal I; x\la \alpha}$. While these
interpretations are not interesting for studying the dynamics of
proof-normalization, they have the virtue of simplicity.

Since Reynolds~\cite{Reynolds}, it is well-known that impredicative or
polymorphic types, as the inhabitants of $\prop$, bear only a trivial
set-theoretical interpretation: if $P:\prop$, then $|P|_{\mathcal I}$ is
either the empty set or a singleton. In other words, all proofs of
proposition $P$ have the same interpretation. Since our theory precisely
identifies all the elements of $P$ at the computational level, the
set-theoretical setting is, for its simplicity the most appealing for
our goal. 

Although the set-theoretical model construction is not as simple as
it might seem~\cite{MiqWer}, the setting is not new; We try to give a
reasonably precise description here.

\subsection{Notations}
Peter Aczel's way to encode set-theoretic functions provides a
tempting framework for a model construction, and a previous version of
this section relied on it. However, because of  technical difficulties
appearing when proving the subject reduction property for the semantic
interpretation we finally favor the traditional set theoretic vision
of functions, where the application $f(x)$ is only defined when $x$
belongs to the domain of the function $f$.

If $\mathcal I$ is a mapping from variables to sets and $\alpha$ is a set,
  we write ${\mathcal   I};x\la \alpha$ for the function mapping $x$ to
  $\alpha$ and identical to ${\mathcal I}$ elsewhere.

The interpretation of the hierarchy $\type(i)$ goes beyond ZFC set
theory and relies on the existence of inaccessible cardinals. This
means, we postulate, for every natural number $n$ the existence of a
set $\U_n$ such that
\begin{enumerate}[$\bullet$]
\item $\U_n\in \U_{n+1}$,
\item $\U_n$ is closed by all set-theoretical operations.
\end{enumerate}

As usual, we write $\emptyset$ for the empty set. We write ${\mathbb
  I}$ for the canonical singleton $\{\emptyset\}$. If $A$ is a set and
  $(B_a)_{a\in A}$ a family of sets indexed over $A$, we use the
  set-theoretical dependent products and sums
\begin{eqnarray*}
 \Pi_{a\in A} B_{a} \eqb \{f\in A\ra \bigcup_{a\in A}B_a~|~\forall
 a\in A.f(a)\in B_a \}\\
 \Sigma_{a\in A} B_{a} \eqb \{(a,b)\in A\times \bigcup_{a\in A}B_a~|~a\in A\wedge b\in B_a\}
\end{eqnarray*}

Finally we write $x\in A\mapsto t$ for the set-theoretical function
construction and, of course, $f(x)$ for set-theoretical function
application.

\subsection{Construction}
Over the ECC fragment of the type theory, the interpretation is
constructed straightforwardly. The fact that non-computational terms
are syntacticly tagged makes the definition easier. We define

\begin{defi}
  For any mapping $\mathcal I$ from variables to
  $\bigcup_{i\in{\mathbb N}}\U_i$, we define a mapping associating a
  set $|\Gamma\vdash t|_{\mathcal I}$ to a term $t$ and a context
  $\Gamma$. This function is defined by induction over the size of $t$
  by the following equations of figure~\ref{semdefr}; we can restrict ourselves to the case
  where $\Gamma\vdash t:T$ for some $T$.
\end{defi}

\begin{figure}

\begin{eqnarray*}
|\Gamma\vdash t|_\I \eqb \emptyset~~~\mbox{if $t$ is of sort $*$}\\
\mbox{In the other cases:}&&\\
|\Gamma\vdash x_\diamond|_\I\eqb \I(x_\diamond)\\
|\Gamma\vdash \lambda x:A.t|_\I\eqb \alpha\in|\Gamma\vdash
A|_\I\mapsto |\Gamma(x:A)\vdash t|_{\I;x\la\alpha}\\
|\Gamma\vdash (t~u)|_\I\eqb |\Gamma\vdash t|_\I(|\Gamma\vdash u|_\I)\\
|\Gamma\vdash \Sigma x:A.B|_{\I}\eqb \Sigma_{\alpha \in
  |\Gamma\vdash A|_{\I}}|\Gamma(x:A)\vdash B|_{\I;x\la\alpha}\\
|\Gamma\vdash \Pi x:A.B|_{\I}\eqb \bigcap_{\alpha \in
  |\Gamma\vdash A|_{\I}}|\Gamma(x:A)\vdash
B|_{\I;x\la\alpha}~~\mbox{if $\Gamma\vdash \Pi x:A.B:\prop$}\\
|\Gamma\vdash \Pi x:A.B|_{\I}\eqb \Pi_{\alpha \in
  |\Gamma\vdash A|_{\I}}|\Gamma(x:A)\vdash
B|_{\I;x\la\alpha}~~\mbox{in the other cases}\\
|\Gamma\vdash <t,u>_{\Sigma x:A.B}|_{\I}\eqb (|\Gamma\vdash
t|_\I,|\Gamma\vdash u|_\I)\\
|\Gamma\vdash \pi_i(t)|_\I\eqb \alpha_i~~~~\mbox{if $|\Gamma\vdash t|_\I$ is a pair
  $(\alpha_1,\alpha_2)$}\\
|\Gamma\vdash \prop|_\I\eqb \{\emptyset ; \II \}\\
|\Gamma\vdash \type(i)|_{\I}\eqb \U_i \\
|\Gamma\vdash (\eqrec~A~P~a~b~p~e)|_\I \eqb |\Gamma\vdash p|_\I
\end{eqnarray*}
\caption{Definition of the semantics}\label{semdefr}

\end{figure}
The following extension of interpretations to contexts is the usual.

\begin{defi}
We define the condition $\I\in |\Gamma|$ by the following clauses
\begin{enumerate}[$\bullet$]
\item $\I\in |[] |$,
\item $\I\in|\Gamma|\wedge \I(x)\in|\Gamma\vdash A|_\I\implies \I\in|\Gamma(x:A)|$.
\end{enumerate}

\end{defi}
This definition should not be surprising. It is a partial definition,
because of two clauses
\begin{enumerate}[$\bullet$]
\item The case of the application, since $|\Gamma\vdash (t~u)|_\I$ is
  only defined when $|\Gamma\vdash t|_\I$ is a (set-theoretic) function and its
  domains contains $|\Gamma\vdash u|_\I$.
\item The cases of the projections, since $|\Gamma\vdash \pi_i(t)|_\I$
  is only defined when $|\Gamma\vdash t|_\I$ is a (set-theoretic) pair.
\end{enumerate}
Note also that the definition depends upon $\Gamma$ only to
discriminate between the case where $\Pi x:A.B$ is impredicative and
is not.
 A more interesting technical point is
the last clause: by anticipating the reduction of $\eqrec$ we have a
total definition which is obviously invariant by reduction.

\begin{lem}[Substitutivity]
Suppose $\Gamma(x_{{\ts}}:U)\Delta\vdash t:T$, $\Gamma\vdash u:U$ and
$\s=s(u)$. Suppose furthermore
\begin{enumerate}[\em(1)]
\item $|\Gamma(x:U) \Delta|$ is defined,
\item if $\I\in|\Gamma(x:U) \Delta| \implies |\Gamma(x:U) \Delta\vdash
  t|_\I \in |\Gamma(x:U) \Delta\vdash T|_\I$,
\item $\I\in|\Gamma|\implies |\Gamma\vdash u|_\I \in |\Gamma\vdash U|_\I$.
\end{enumerate}

Let $\I\in |\Gamma\Delta[x\setminus u]|$; we have
\begin{enumerate}[$\bullet$]
\item  $|\Gamma\Delta[x\setminus u]\vdash t[x\setminus u]|_\I$ is
defined and equal to 
$ |\Gamma(x_{{\ts}}:U)\Delta\vdash t |_{\I;x\la |\Gamma\vdash u|_\I}$
\item $(\I;x\la |\Gamma\vdash u|_\I)\in \|\Gamma(x:U) \Delta|$.
\end{enumerate}
\end{lem}
\begin{proof}
By a simple induction over the structure of the derivation.

Note that in the case where $t$ is of the form $\Pi y:A.B$, one uses
the fact that typing is preserved by substitution
(lemma~\ref{wtsubst}) in order to ensure that the applied clause
remains the same ($\Pi x:A.B$ being of type $\prop$ or $\type(i)$).
\end{proof}

\begin{lem}[Correctness for reduction]
Let  $\Gamma\vdash t:T$ be derivable; have $\I\in|\Gamma|$ such that
$|\Gamma\vdash t|_\I$ is defined. If 
$t\rhd_\varepsilon t'$, then $|\Gamma\vdash t'|_\I = |\Gamma\vdash t|_\I$.
\end{lem}
\begin{proof}
By induction over the typing derivation. As pointed out
in~\cite{MiqWer}, the restriction on the $\beta$-reduction that ensures
that the tag does not change is essential here.
\end{proof}
\begin{cor}
Let  $\Gamma\vdash t:T$ and $\Gamma\vdash t':T$ be derivable; have
$\I\in|\Gamma|$ such that 
$|\Gamma\vdash t|_\I$ and $|\Gamma\vdash t'|_\I$ are defined.
 $t\conv t'$, then we have $|t|_\I = |t'|_\I$.
\end{cor}

Soundness is then proved without much difficulty.

\begin{thm}
If $\Gamma\wf$ is derivable, then $|\Gamma|$ is defined. 
If $\Gamma\vdash t:T$ is derivable,  and $\I\in|\Gamma|$ then
 $|\Gamma\vdash t|_\I\in|\Gamma\vdash T|_\I$ (and both objects are defined).
\end{thm}
\proof
By induction over the
derivation. When checking the correctness of the interpretation of
$\eqrec$, one simply has to remark that propositional Leibniz equality
is indeed interpreted by set-theoretical equality; that is, if
$|\Gamma\vdash a|_\I$ and $|\Gamma\vdash b|_\I$ are both elements of $|\Gamma\vdash A|_\I$, then 
\begin{enumerate}[$\bullet$]
\item $|\Gamma\vdash a =_A b|_\I = \II$ if $|\Gamma\vdash a|_\I = |\Gamma\vdash b|_\I$,
\item $|\Gamma\vdash a =_A b|_\I = \emptyset$ if $\Gamma\vdash |a|_\I \neq |\Gamma\vdash b|_\I$.\qed
\end{enumerate}

It is easy to check that the axioms AC, EM and EXT of the previous
section are valid in this model.

\section{Conclusion and further work}
\noindent We have tried to show that a relaxed conversion rule can make type
theories more practical, without necessarily giving up normalization
or decidable type checking. In particular, we have shown that this
approach brings closer the world of PVS and type theories of the Coq
family.

We also view this as a contribution to closing the gap between proof
systems like Coq and safe programming environments like Dependent ML
or ATS~\cite{Xi1,Xi2}. But this will only be assessed by practice; the
first step is thus to implement such a theory.

\section{Acknowledgements}
\noindent Christine Paulin deserves special thanks, since she me gave
the idea of a relaxed conversion rule long ago as well as the hint for
the computation rule of paragraph~\ref{eqrec}. Bruno Barras was very
helpful by pointing out some errors and Makoto Tatsuta by helping me
to check that defining simultaneously reduction and conversion did not
break Church-Rosser. A question of Russell O'Connor suggested the
generalization of paragraph~\ref{ROC}. This work also benefited from
discussions with Bruno Barras, Hugo Herbelin and Benjamin Grégoire.
Anonymous referees gave numerous useful comments and hints.

\end{document}